# Force-gradient-induced mechanical dissipation of quartz tuning fork force sensors used in atomic force microscopy

A Castellanos-Gomez[1], N Agraït[1,2,3] and G Rubio-Bollinger[1,2]

[1] Departamento de Física de la Materia Condensada (C–III).
Universidad Autónoma de Madrid, Campus de Cantoblanco, 28049 Madrid, Spain.
[2] Instituto Universitario de Ciencia de Materiales "Nicolás Cabrera".
Universidad Autónoma de Madrid, Campus de Cantoblanco, 28049 Madrid, Spain.
[3] Instituto Madrileño de Estudios Avanzados en Nanociencia
IMDEA-Nanociencia, 28049 Madrid, Spain.

E-mail: andres.castellanos@uam.es  and  gabino.rubio@uam.es
+34 91 497 5552
+34 91 497 3961
C\ Fco. Tomas y Valiente, 7.
Facultad de Ciencias.
Dpto. Física de la Materia Condensada (C-III).
Lab. 201.
C.P. 28049. (Madrid).
SPAIN.

We have studied the dynamics of quartz tuning fork resonators used in atomic force microscopy taking into account mechanical energy dissipation through the attachment of the tuning fork base. We find that the tuning fork resonator quality factor changes even for the case of a purely elastic sensor-sample interaction. This is due to the effective mechanical imbalance of the tuning fork prongs induced by the sensor-sample force gradient which in turn has an impact on the dissipation through the attachment of the resonator base. This effect may yield a measured dissipation signal that can be different to the one exclusively related to the dissipation between the sensor and the sample. We also find that there is a second order term in addition to the linear relationship between the sensor-sample force gradient and the resonance frequency shift of the tuning fork that is significant even for force gradients usually present in atomic force microscopy which are in the range of tens of N/m.

## 1. Introduction

Miniaturized quartz tuning forks are piezoelectric crystal resonators used as time-bases in the watch and electronic industries. These quartz tuning forks have been modified to be used as force sensors in scanning probe microscopes by attaching a sharp tip to the free end of one of the prongs [1]. These force sensors have some advantages over conventional force sensors, based on microfabricated cantilevers, such as their self-sensing piezoelectric current read out, higher quality factor Q and higher stiffness which prevents the tip jumping to the contact. These characteristics have made atomic resolution imaging possible [2, 3] and high sensitivity measurement of atomic scale forces [4, 5]. In addition, another interesting feature is that these sensors can be used in a variety of environments such as ultrahigh vacuum, low temperature and high magnetic field. Traditionally two different working schemes have been used with tuning fork sensors. In the first one the tuning fork is used with the two prongs oscillating freely [6]. In the second one the tipless prong is firmly glued to a massive holder [7] turning the tuning fork into a quartz cantilever (qPlus scheme). The qPlus sensors have the main advantages of having a cantilever-like dynamics and a lower effective elastic constant keff. However tuning fork sensors can have much higher Q factor and produce at least twice the piezoelectric current of a qPlus sensor for a given tip oscillation amplitude. Fig. 1 shows a comparison of the resonance spectra measured for a qPlus sensor (blue squares) and a tuning fork (red circles) installed in a dilution fridge at a temperature of 7 mK. The ultrahigh Q factor of the tuning fork sensor is one order of magnitude larger than the one obtained with the qPlus sensor. In order to use tuning fork sensors and take advantage of their higher Q factor and higher piezoelectric signal, a detailed analysis of the tuning fork dynamics has been performed.

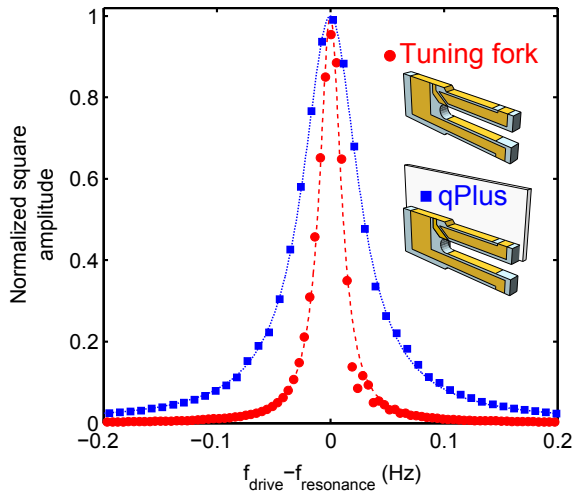

**Fig. 1.** Resonance spectra of a balanced tuning fork (red circles) and a qPlus sensor (blue squares) measured at a temperature of $T$ = 7 mK. The resonance frequency of the tuning fork (qPlus) is 32710.5 Hz (12492 Hz) and its $Q$ factor is $1.4 \cdot 10^6$ ($2.1 \cdot 10^5$). A fit to a Lorentzian curve is also shown (lines).

The dynamics of tuning fork sensors imbalanced by the weight of the tip has been already studied [8-11]. In this work we have carried out a study of the changes that influences the tuning fork dynamics induced by the typical force gradients acting on the tip during atomic force microscopy (AFM) experiments. We have used a simplified mechanical model to calculate the dynamics of force gradient induced imbalance of tuning forks. The remarkable agreement between the calculated dynamics and the one measured for mass loaded tuning forks validates the model. We find that the force gradient induced imbalance changes the tuning fork Q factor even for the case of purely elastic tip-sample interaction. We also find that for very precise measurements of force gradients the relationship between the force gradient and the frequency shift can not be considered linear but a non negligible second order term must be taken into account even, for force gradients usually present in atomic force microscopy.

## 2. Tuning fork dynamical model

Tuning fork dynamics have been widely modelled as a single harmonic oscillator [1, 12-14], where the tuning fork prongs behave like uncoupled cantilevers, but the experimentally observed dynamics can be more accurately explained using a coupled harmonic oscillator model [8, 11]. Therefore, the coupling between the tuning fork prongs plays an important role in the tuning fork dynamics. Indeed it has been shown that a simplified coupled oscillator model can account for the basic tuning fork dynamical behavior in a remarkable way [15]. However, the absence of damping elements in the model does not permit the explanation of (at least) two experimental observations. First, the *Q* factor of the tuning fork is drastically reduced when the tuning fork is imbalanced by mass loading one of the prongs. Second, when the tuning fork is excited mechanically (vibrating its base at a given amplitude), the resulting oscillation amplitude is larger for imbalanced tuning forks even though the *Q* factor is lower than for balanced tuning forks.

We have used a coupled oscillator model with two damping terms [8]. The first one accounts for the mechanical energy dissipation of each prong with a damping coefficient $\gamma$ such as that arising from hydrodynamic friction. Second, the dissipation through the attachment of the tuning fork base has been effectively taken into account adding a term $\gamma_{\text{base}}$ proportional to the velocity of the center of mass of the tuning fork. The motion of the masses 1 and 2 ($x_1$ and $x_2$) corresponds to the deflection of the free end of the prongs 1 and 2 of the tuning fork. The equations of motion for the masses are:

$$\left. \begin{array}{l} \alpha m \ddot{x}_1 + \beta k \cdot x_1 + k_c (x_1 - x_2) + \dfrac{\gamma_{\text{base}} m}{1+\alpha}(\alpha \dot{x}_1 + \dot{x}_2) + \gamma m \cdot \dot{x}_1 = F_1(t) \\ m \ddot{x}_2 + k \cdot x_2 + k_c (x_2 - x_1) + \dfrac{\gamma_{\text{base}} m}{1+\alpha}(\alpha \dot{x}_1 + \dot{x}_2) + \gamma m \cdot \dot{x}_2 = F_2(t) \end{array} \right\}, \qquad (1)$$

where *k* is the elastic constant of one prong, $k_c$ is the elastic constant of the coupling between the prongs and *m* is the effective mass of one prong. The parameter $\alpha = 1 + \Delta m / m$ takes into account the effect of an extra mass $\Delta m$ if a tip is attached to prong 1, the parameter $\beta = 1 + (\partial F / \partial z)/k$ takes into account the effect of an external force gradient ($\partial F / \partial z$) acting on the tip and $F_i(t)$ is the driving force on mass *i*=1,2. Setting $\beta \cdot F_1(t) = F_2(t) = F_0 \cdot \cos(\omega t)$ we simulate the mechanical excitation of the tuning fork at frequency $\omega$. Note that this is different than electrical self-excitation which would enter the model by setting $F_1(t) = -F_2(t) = F_0 \cdot \cos(\omega t)$ resulting in a different dynamics which will not be addressed here. There are two eigenmodes solutions of equations 1: one in which both masses oscillate in phase and another with the masses oscillating in anti-phase (in-phase and anti-phase modes hereafter). The measured frequencies of these eigenmodes, of the balanced tuning fork under study1, can be used to unambiguously determine the values of *k* and $k_c$. Subsequently, the measured *Q* factor of these eigenmodes can be used to find the value of the $\gamma$ and $\gamma_{\text{base}}$ terms. All these parameters of the model remain unaltered when the tuning fork is unbalanced either by a force gradient acting on a prong or by a change of the mass load. In this way the value of the parameters which determine the dynamics of unbalanced tuning forks are experimentally obtained for a given balanced tuning fork and base attachment implementation. An outcome of the model is that the change in the tuning fork oscillation produced by mass loading ($\Delta m$) the tuning fork prong 1 is equivalent, to a first order approximation, to an external negative force gradient ($\partial F / \partial z$) acting on the prong:

$$\frac{\partial F}{\partial z} = -k \left(\frac{f_{20}}{f_{10}}\right)^2 \frac{\Delta m}{m} = -4\pi^2 f_{20}^{\;2} \cdot \Delta m, \qquad (2)$$

where $f_{10}$ and $f_{20}$ are the frequencies of the in-phase and anti-phase modes of the balanced tuning fork. Both frequencies can be measured as described in ref. [15] and in the supplemental material. Equation (2) shows that a positive force gradient acting on prong 1 can be used to rebalance a tip loaded tuning fork. A similar strategy has been used in previous works to rebalance a tuning fork by increasing the prong stiffness [16, 17]. As a consequence of the equivalence between mass load and force gradient, the response of tuning fork sensors to mass loads can be used to prove the validity of the coupled oscillators model. Note that applying a well defined force gradient to a tuning fork prong can be experimentally challenging while adding test masses at the end of one prong can be routinely achieved [18].

---

[1] The tuning forks are purchased from Digikey, part number SE3301-ND, Epson Toyocom C-2 Series 20 KHz, $k = 980$ N/m. For balanced tuning forks the eigenfrequencies are $f_{\text{in-phase}} = 18.5$ kHz and $f_{\text{anti-phase}} = 20$ kHz and their *Q* factors are $Q_{\text{in-phase}} \sim 950$ and $Q_{\text{anti-phase}} \sim 4350$.

## 3. Results

We have used a setup similar to the one described in detail in ref. [15] to study the dynamics of mass loaded tuning forks under ambient conditions. A dither piezoelectric element mechanically excites the tuning fork oscillation while it is inspected under a Nikon Eclipse LV-100 optical microscope and the piezoelectric current of the tuning fork is measured. It is worth noting that we have observed that the small oscillation linear regime is valid for amplitudes up to tens of microns [15]. This enables the accurate measurement of the oscillation amplitude of the prongs by inspection under an optical microscope. Using stroboscopic illumination at twice the driving frequency facilitates the measurement of the oscillation amplitude given by the distance between the two turning points of the oscillation [15]. A digital camera is used to acquire the image of the oscillating tuning fork, which appears freezed under stroboscopic illumination. The digital image, whose calibration is 45 nm per pixel, is analyzed to determine the oscillation amplitude with accuracy better than 1%. This technique is used to follow the change of the oscillation amplitude of each prong when the tuning fork is unbalanced, keeping the driving amplitude fixed. The tuning forks have been mass loaded by attaching test masses at the end of one of their prongs. The employed test masses are lead spheres (15-45 μm diameter) whose mass is determined by measuring their diameter with the optical microscope and using the density of the bulk material [18].

Fig. 2(a) shows the measured resonance spectra for a mass loaded tuning fork excited in the anti-phase mode and the corresponding resonance frequency shift is shown in Fig. 2(b). Fig. 2(c) shows the oscillation amplitude ratio of the tuning fork prongs $\left(A_{prong1} / A_{prong2}\right)$ that is unity for a perfectly balanced tuning fork and decreases with the mass load of prong 1. When the oscillation amplitude of each prong is different the center of mass of the whole tuning fork moves causing a finite mechanical dissipation through the attachment of the tuning fork base and consequently a reduction of the $Q$ factor (Fig. 2(d)).

On the other hand the tuning fork imbalance also affects the effectiveness of the external mechanical excitation of the tuning fork oscillation. As shown in Fig. 2(e), for a fixed mechanical excitation amplitude, the stronger the imbalance the more effective the mechanical excitation will be. This is because an increasing difference between the anti-phase oscillation amplitude of each prong is related to an increase of the tuning fork center of mass oscillation amplitude, (and reciprocally)??. In addition, the resulting oscillation amplitude is also modulated by the change of the $Q$ factor acting as a competing effect. For the studied tuning fork the net result is an increase of the summed oscillation amplitude $\left( A_{prong1} + A_{prong2} \right)$ with the mass load, but this tendency depends on details of the tuning fork arrangement. The tuning fork dynamics characterization is compared in Fig. 2(b)-(e) to the result of the coupled oscillator model (dashed lines) where the driving force $F_0$ is a parameter adjusted to match the slope of the measured $\left( A_{prong1} + A_{prong2} \right)$ vs. mass load curve (Fig. 2(e)).

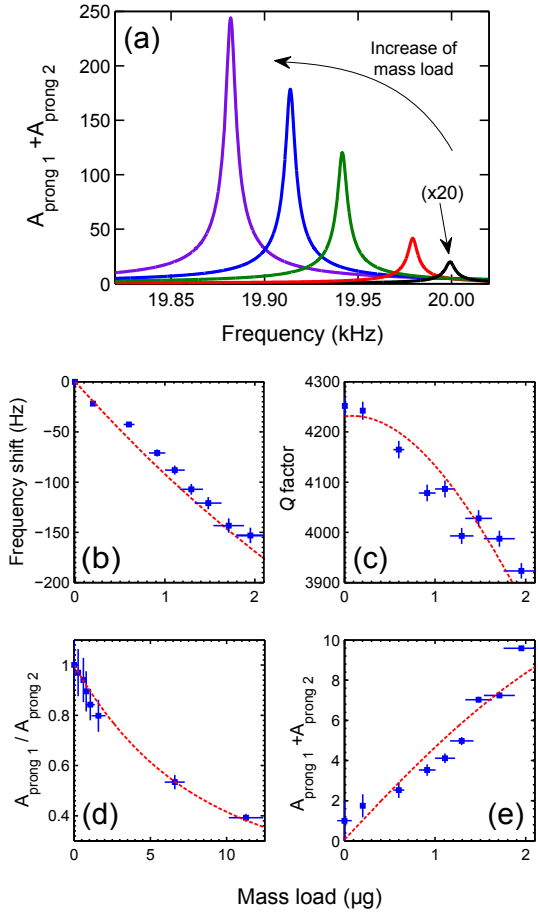

**Fig. 2.** (a) Resonance spectra of a tuning fork with an increasing mass load on prong 1 (from right to left). (b-e) Comparison between the measured (symbols) and calculated (lines) frequency shift, prong's amplitude ratio, the summed oscillation amplitude $\left( A_{prong1} + A_{prong2} \right)$ and the $Q$ factor. The summed oscillation amplitude has been normalized to the amplitude of the unloaded tuning fork.

The model can be used to calculate the first order linear relationship between the frequency shift and the force gradient acting on the tip. The proportionality constant involves an effective elastic constant $k_{eff}$ which depends on the initial imbalance of the tuning fork due to the presence of a tip attached at the end of one of the prongs [8-11]. While it is sufficient to know the mass load of the tip ($\Delta m/m$), which can be obtained from its geometry and density, the effective elastic constant can be also obtained using equation (1) from the measurement of the frequency of the anti-phase and in-phase modes ($f_{10}$ and $f_{20}$ respectively) before the tip is mounted on the prong and the frequency of the anti-phase mode ($f$) after the tip is mounted:

$$\frac{\Delta m}{m} = 2\left[\frac{(f^2 - f_{10}^2)(f^2 - f_{20}^2)}{f^2(f_{10}^2 + f_{20}^2 - 2f^2)}\right]. \tag{3}$$

Once the tip mass load $\Delta m / m$ has been determined, the resonance frequency shift and the force gradient are related by $\frac{\Delta f}{f} \simeq \frac{1}{2}\frac{\partial F/\partial z}{k_{eff}}$ and $k_{eff}$ is obtained, to a first order approximation, from the solution of the eigenvalue equation (1):

$$k_{eff} = -\frac{k}{8\pi^2 f_{10}^2} \cdot \frac{B(A+B)}{(C-B)}, \tag{4}$$

where

$$\begin{aligned} A &= 4\pi^2 \cdot (\alpha+1) \cdot (f_{10}^2 + f_{20}^2), \\ B &= \sqrt{A^2 - 256\pi^4 \alpha f_{10}^2 f_{20}^2}, \\ C &= \left(\frac{\alpha-1}{\alpha+1}\right) A. \end{aligned} \tag{5}$$

The elastic constant $k$ of the tuning fork prong can be obtained from its geometry. We use equations (3), (4) and (5) to determine the effective elastic constants of tuning fork sensors supplemented with commonly used tips. For instance, in the case of a carbon fiber tip [19] (250 μm long and 7 μm in diameter) $k_{eff}$ = 2650 N/m while for a platinum iridium tip [20] (250 μm long and 25 μm in diameter) $k_{eff}$ = 24500 N/m which is one order of magnitude higher. Therefore light tips are superior with respect to sensitivity and special attention has to be paid to avoid excessive mass loading of the tuning fork prongs which drastically reduces the tuning fork sensitivity to the tip-sample interaction.

The dynamical model allows evaluating the weight of second order corrections to the relationship between the frequency shift and the force gradient, which will be shown to be relevant for commonly encountered force gradients in AFM experiments (in a range -20 to 20 N/m) in non-contact AFM [21] and nanoindentation measurements [4, 5]. Up to second order the frequency shift is given by:

$$\frac{\Delta f}{f} \simeq \frac{1}{2k_{eff}} \cdot \frac{\partial F}{\partial z} + \delta \cdot \left(\frac{\partial F}{\partial z}\right)^2 + O\left(\left(\frac{\partial F}{\partial z}\right)^3\right). \tag{6}$$

It is straightforward to obtain the value of $\delta$ however the expression looks somewhat cumbersome even for the simplified coupled oscillators model:

$$\delta = \frac{16\pi^4 f_{10}^4}{k^2 B^2 (A+B)} \left[\frac{(B^2 - C^2)}{B} - \frac{(C-B)^2}{A+B}\right]. \tag{7}$$

We have found that this second order term can be 30 times larger for tip loaded tuning forks than for qPlus sensors. In Fig. 3(a) we show the numerically calculated non-linear relationship between the frequency shift and the force gradient for a carbon fiber tip supplemented tuning fork [19]. The significant deviation from the linear regime puts forward the importance of the second order term even for small tip-sample force gradients.

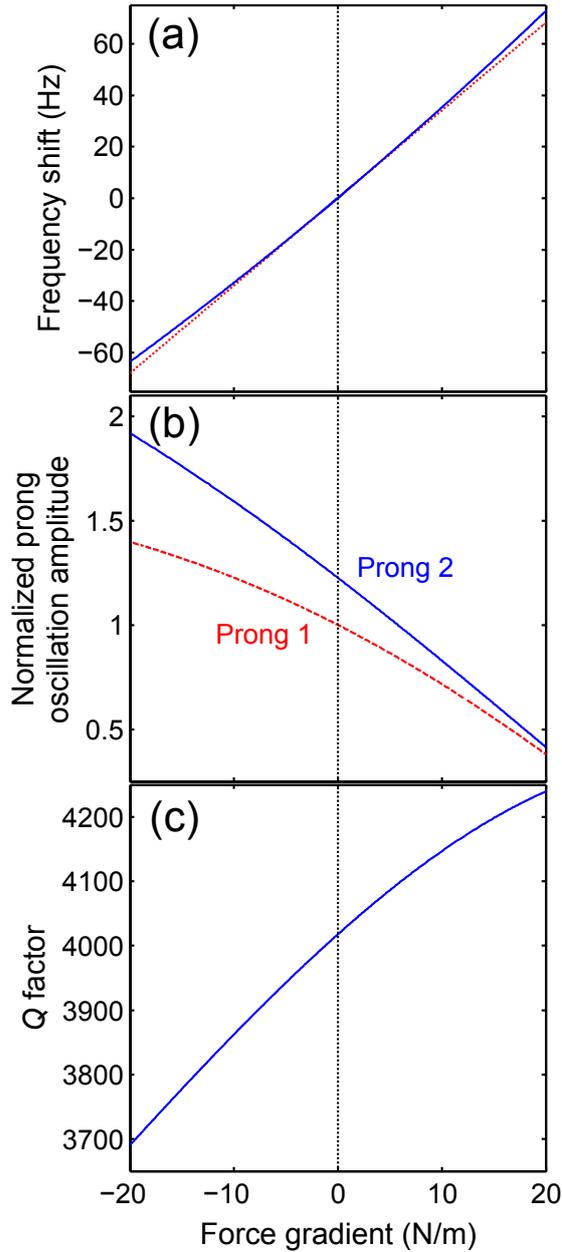

**Fig. 3.** Calculated frequency shift (a), prongs oscillation amplitudes (b) and *Q* factor (c) as a function of the force gradient applied to the tuning fork tip. The frequency shift *vs.* force gradient curve shows a noticeable deviation from the linear relationship (dotted line in Fig. 3(a)). Prongs oscillation amplitudes are normalized to the value of the prong 1 oscillation amplitude when no force gradient is applied.

Fig. 3(b) shows how an external force gradient applied to one of the prongs changes the oscillation amplitude of each of the prongs in a different manner. This dynamical behavior affects the apparent dissipation in two ways. First, a purely elastic sensor-sample interaction causes a change to the mechanical excitation effectiveness and therefore the corresponding change of the oscillation amplitude which can be erroneously interpreted as dissipative losses at the tip-sample interface. Second, the external force gradient induced imbalance changes the energy losses through the tuning fork base attachment. This is a real dissipation channel but it is unrelated to the dissipation at the tip-sample interface. The relevance of the change of the excitation effectiveness is quantified in Fig. 3(b). The summed oscillation amplitude, and thus the measured piezoelectric current, drops 60% for 20 N/m positive force gradient while it increases 50% for -20 N/m force gradient. On the other hand Fig. 3(c). quantifies the effect of the imbalance induced energy losses, exclusively through the tuning fork attachment, with its corresponding impact on the *Q* factor of the resonator which changes by 5%. While the effect of the change of the excitation effectiveness and the energy losses through the attachment as a function of the force gradient have dependencies of opposite sign they will rarely compensate each other, especially because of the variability of the base attachment quality and strength which is usually mediated by glue.

Although all the above described effects have to be carefully evaluated for quantitative measurements of force gradient and dissipation when an AFM is used to obtain images at constant force gradient, dissipation changes are related to real dissipation at the tip-sample interface. The situation turns out to be more delicate if scanning is not performed at constant frequency shift, that is, when it is not operated in the frequency modulation AFM mode (FM-AFM). That would be the case when amplitude modulation AFM mode (AM-AFM) is used as keeping the amplitude fixed does not ensures a fixed value for the force gradient. In addition a detailed analysis is required for a quantitative interpretation of force gradient *vs.* tip-sample distance experiments because of the interplay between a purely elastic force gradient, a variable unbalancing of the tuning fork and the unavoidable losses through the tuning fork base attachment. These effects are also expected to be present not only under ambient conditions but also in other environments such as ultrahigh vacuum or low temperature. In such situations one of the main differences would be a dramatic lowering of the hydrodynamic or intrinsic damping term ($\gamma$). According to the dynamical model, the non-linear relationship between the force gradient and the frequency shift would remain unaltered since it does not depend on the damping terms ($\gamma$ and $\gamma_{base}$). On the contrary, low values of the damping of the prongs ($\gamma$) would enhance the $Q$ factor of the oscillation. Therefore, the impact on the reduction of the $Q$ factor related to the force gradient induced imbalance, due to losses through the base attachment, would be highly noticeable. Although a high $Q$ factor is desirable, in principle, to detect low dissipation and force gradient, the impact of this adverse effect would make it difficult to disentangle the dissipation losses at the tip-sample interface from those through the tuning fork base attachment, because of the increased relative strength of the $\gamma_{base}$ term in the dynamical equations. It has to be stressed that, despite the qPlus tuning fork configuration being usually associated with a lower $Q$ factor, the resonance frequency shift and the dissipation changes are straightforward to interpret in terms of the physical phenomena occurring solely between the tip and the sample [7].

## 4. Conclusions

We have used a simplified mechanical model to study of the dynamics of a tuning fork, used in atomic force microscopy, which includes dissipation mechanisms of the hydrodynamic damping of the oscillating prongs and energy losses through the attachment of the tuning fork base. We find that the effective elastic constant which relates the frequency shift of the resonator to the force gradient acting on the tip strongly increases with mass load due to the tip attachment and causes a dramatic drop of the force gradient sensitivity. In addition for realistic tip mass loads the second order correction in the relationship between the frequency shift and the force gradient is not negligible at force gradients commonly present in atomic force microscopy experiments. On the other hand a purely conservative tip-sample interaction causes the imbalance of the tuning fork. The consequences of this imbalance in the apparent dissipation are twofold. First, the mechanical excitation effectiveness, and thus the oscillation amplitude, changes with the force gradient induced imbalance and this can be misinterpreted as a dissipative tip-sample interaction. Second, there is a real dissipation, unrelated to the tip-sample interaction, through the tuning fork base attachment which is also dependent on the force gradient induced imbalance. Therefore a detailed analysis of the tuning fork dynamics is required for a quantitative interpretation of frequency shift and dissipation measurements in terms of the physical phenomena occurring exclusively between the tip and the sample.

## 5. Acknowledgements

A.C-G. acknowledges fellowship support from the Comunidad de Madrid (Spain). This work was supported by MICINN (Spain) (MAT2008-01735 and CONSOLIDER en Nanociencia molecular CSD-2007-00010).

# Supplemental information:
## Force-gradient-induced mechanical dissipation of quartz tuning fork force sensors used in atomic force microscopy

A Castellanos-Gomez, N Agraït and G Rubio-Bollinger

**1. Eigenfrequencies for high *Q* tuning fork sensors imbalanced by an arbitrary mass load and force gradient.**

For small damping constants $\gamma$ and $\gamma_{base}$, which can always be assumed in the case of high $Q$ resonators, the eigenfrequencies of the in-phase and the anti-phase modes ($\omega_1$ and $\omega_2$ respectively) can be obtained by solving the eigenvalue problem of expression (1) in the manuscript neglecting the damping and forcing terms [1]:

$$\omega_{2,1}^2 \simeq \frac{1}{4\alpha}\left\{\Omega \pm \sqrt{\Omega^2 - 8\alpha\omega_{10}^2\left[\omega_{10}^2(\beta-1)+\omega_{20}^2(\beta+1)\right]}\right\} \quad . \tag{S8}$$

where $\Omega = \left[\omega_{10}^2(\alpha+2\beta-1)+\omega_{20}^2(\alpha+1)\right]$, the parameter $\alpha = 1+\Delta m/m$ takes into account the effect of an extra mass $\Delta m$ if a tip is attached to prong 1, the parameter $\beta = 1+(\partial F/\partial z)/k$ takes into account the effect of an external force gradient $\partial F/\partial z$ acting on prong 1 and $k$ and $m$ are respectively the elastic constant and the effective mass of a tuning fork prong. The terms $\omega_{10}$ and $\omega_{20}$ are the eigenfrequencies of the in-phase and anti-phase modes for a perfectly balanced tuning fork ($\alpha=1$ and $\beta=1$). These eigenfrequencies can be expressed in terms of the elastic constant of a prong $k$, the effective mass of a prong $m$ and the elastic constant of the coupling between prongs $k_c$:

$$\begin{aligned}\omega_{10} &= 2\pi f_{10} = \sqrt{\frac{k}{m}}, \\ \omega_{20} &= 2\pi f_{20} = \sqrt{\frac{k+2k_c}{m}}.\end{aligned} \tag{S9}$$

Although the electrodes configuration in commercially available quartz tuning fork resonators is designed to null out the piezoelectric current from the in-phase mode, in practice, it is possible to determine the resonance frequency of both the in-phase and the anti-phase modes from electric measurements [2] due to small imperfections in the electrodes design (figure S1).

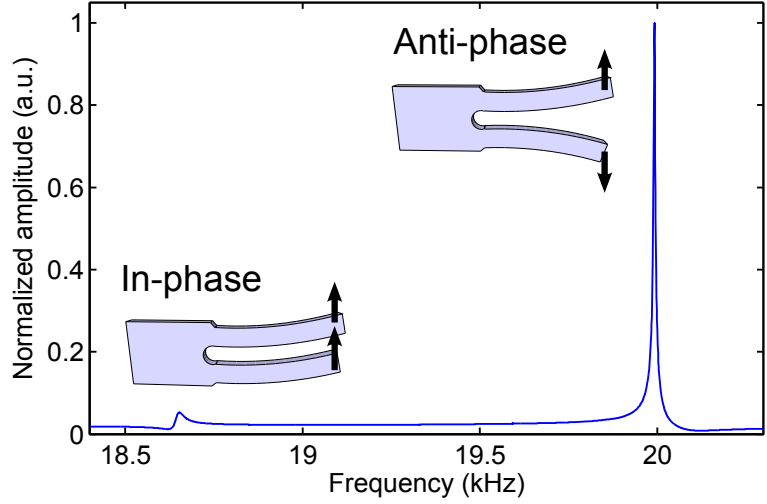

**Figure S4:** Resonance spectrum of a quartz tuning fork (Epson Toyocom C-2 Series 20 KHz) obtained by measuring the piezoelectric current wile the tuning fork is excited mechanically. Both the in-phase and the anti-phase modes can be observed. Optical inspection has been used to check correspondence of these peaks with the in-phase and the anti-phase modes.

## 2. Effect of small $\Delta m$ and $\partial F / \partial z$ on the tuning fork dynamics.

Expression (S1) can be simplified in the case of a tuning fork slightly imbalanced by a low mass load $(\alpha \to 1)$ and a low force gradient $(\beta \to 1)$ acting on prong 1. The eigenfrequencies are, at first order, obtained from the Taylor series expansion of eq. S1:

$$\omega_1^2 \simeq \omega_{10}^2 \cdot \left(1 + \frac{\partial F / \partial z}{2k} - \frac{\Delta m}{2m} + ...\right),$$
$$\omega_2^2 \simeq \omega_{20}^2 \cdot \left[1 + \left(\frac{\omega_{10}}{\omega_{20}}\right)^2 \cdot \frac{\partial F / \partial z}{2k} - \frac{\Delta m}{2m} + ...\right].$$
(S10)

Using $\sqrt{1 + a\mathrm{x} + b\mathrm{y}} \xrightarrow[y \to 0]{x \to 0} 1 + \frac{a}{2}\mathrm{x} + \frac{b}{2}\mathrm{y} + ...$ in expression S3 we obtain:

$$\frac{\omega_1}{\omega_{10}} \simeq \sqrt{1 + \frac{\partial F / \partial z}{2k} - \frac{\Delta m}{2m} + ...} \simeq 1 + \frac{\partial F / \partial z}{4k} - \frac{\Delta m}{4m} + ... ,$$
$$\frac{\omega_2}{\omega_{20}} \simeq \sqrt{1 + \left(\frac{\omega_{10}}{\omega_{20}}\right)^2 \cdot \frac{\partial F / \partial z}{2k} - \frac{\Delta m}{2m} + ...} \simeq 1 + \left(\frac{\omega_{10}}{\omega_{20}}\right)^2 \cdot \frac{\partial F / \partial z}{4k} - \frac{\Delta m}{4m} + ... .$$
(S11)

Therefore, the change in the dynamics of the anti-phase mode produced by a mass load $\Delta m$ is equivalent to the change produced by a certain force gradient $\partial F / \partial z$ given by:

$$\frac{\partial F}{\partial z} = -\left(\frac{\omega_{20}}{\omega_{10}}\right)^2 \frac{k}{m} \cdot \Delta m = -\omega_{20}^2 \cdot \Delta m \cdot$$
(S12)

Which gives expression (2) in the manuscript.

## 3. Determination of the mass load $\Delta m$ from the change produced on the tuning fork dynamics.

Considering that the tuning fork prong 1 is imbalanced by a mass load $(\alpha = 1 + \Delta m/m)$ and that there is no force gradient acting on that prong $(\beta = 1)$, expression S2 can be written as:

$$\omega_{2,1}^2 \simeq \frac{1}{4\alpha}\left(A \pm \sqrt{A^2 - 16\alpha\omega_{10}^2\omega_{20}^2}\right), \tag{S13}$$

where $A = (\alpha + 1) \cdot (\omega_{10}^2 + \omega_{20}^2)$. Solving for $\alpha$ in eq. S6 and using $\alpha = 1 + \Delta m/m$, the relationship between the mass load $\Delta m/m$ and the change in the resonance frequency is obtained:

$$\frac{\Delta m}{m} = 2\left[\frac{(\omega^2 - \omega_{10}^2)(\omega^2 - \omega_{20}^2)}{\omega^2(\omega_{10}^2 + \omega_{20}^2 - 2\omega^2)}\right], \tag{S14}$$

Which is expression (3) in the manuscript.

## 4. Relationship between the force gradient and the resonance frequency shift for tuning fork resonators imbalanced by the tip mass load $\Delta m$.

Considering that the tuning fork prong 1 is imbalanced by the tip mass load $\Delta m$ and a small force gradient $(\beta = 1 + \partial F/\partial z \to 1)$, expression S2 can be expanded in Taylor series as follows:

$$\omega_{2,1}^2 \simeq \frac{1}{4\alpha}(A \pm B) \mp (\beta - 1) \cdot (C \mp B) \cdot \frac{\omega_{10}^2}{2\alpha B} \pm (\beta - 1)^2 \cdot (B^2 - C^2) \cdot \frac{\omega_{10}^4}{2\alpha B^3} + \ldots, \tag{S15}$$

where $B = \sqrt{A^2 - 16\alpha\omega_{10}^2\omega_{20}^2}$ and $C = (\alpha - 1/\alpha + 1)A$. Defining $\omega_{2,1}(\beta = 1) \equiv (A \pm B)/4\alpha$ and rearranging expression S8:

$$\frac{\omega_{2,1}}{\omega_{2,1}(\beta = 1)} \simeq \sqrt{1 \mp (\beta - 1) \cdot \left(\frac{C \mp B}{A \pm B}\right) \cdot \frac{2\omega_{10}^2}{B} \pm (\beta - 1)^2 \cdot \left(\frac{B^2 - C^2}{A \pm B}\right) \cdot \frac{2\omega_{10}^4}{B^3} + \ldots}, \tag{S16}$$

and using the Taylor series expansion $\sqrt{1 + ax + bx^2} \xrightarrow{x \to 0} 1 + \frac{a}{2}x + \left(\frac{b}{2} - \frac{a^2}{8}\right)x^2 + \ldots$, expression S9 can be approximated as:

$$\begin{aligned}\frac{\Delta\omega_{2,1}}{\omega_{2,1}} &\equiv \frac{\omega_{2,1} - \omega_{2,1}(\beta = 1)}{\omega_{2,1}(\beta = 1)} \simeq \ldots \\ &\simeq \mp(\beta - 1) \cdot \left(\frac{C \mp B}{A \pm B}\right) \cdot \frac{\omega_{10}^2}{B} \pm (\beta - 1)^2 \cdot \left[\frac{B^2 - C^2}{B} - \frac{(C \mp B)^2}{A \pm B}\right] \cdot \frac{\omega_{10}^4}{B^2(A \pm B)} + \ldots\end{aligned} \tag{S17}$$

Terms in S10 can be identified as expressions (4), (6) and (7) in the manuscript.